\documentclass[aps,twocolumn,showpacs]{revtex4}
\usepackage{amsmath,amssymb}
\usepackage{graphicx}

\def\v#1{\mathbf #1}
\def\dfrac#1#2{{\displaystyle\frac{#1}{#2}}}
\newcommand{\bra}[1]{\left\langle {#1} \right\vert}
\newcommand{\ket}[1]{\mid\!{#1}\,\rangle}
\newcommand{\kket}[1]{\parallel\!{#1}\,\rangle\!\rangle}
\newcommand{\braket}[2]{\left\langle {#1} \right\vert\left. {#2}\right\rangle}
\newcommand{\aver}[1]{\langle {#1}\rangle}

\begin{document}

\title{Disordered ground states in a quantum frustrated spin chain 
with side chains}

\author{Ken'ichi Takano$^{1}$ and Kazuo Hida$^{2}$}

\affiliation{$^{1}$Toyota Technological Institute, 
Tenpaku-ku, Nagoya 468-8511, Japan \\
$^{2}$Division of Material Science, 
Graduate School of Science and Engineering, \\ 
Saitama University, Saitama 338-8570, Japan}

\date{\today}

\begin{abstract}
We study a frustrated mixed spin chain with side chains, 
where the spin species and the exchange interactions are 
spatially varied. 
A nonlinear $\sigma$ model method is formulated for this model, 
and a phase diagram with two disordered spin-gap phases 
is obtained for typical cases. 
Among them we examine the case with a main chain consisting of 
an alternating array of spin-1 and spin-$\frac{1}{2}$ sites 
and  side chains each of a single spin-$\frac{1}{2}$ site 
in great detail.  
Based on  numerical, perturbational, and variational approaches, 
we propose a singlet cluster solid picture for each phase, 
where the ground state is expressed as 
a tensor product of local singlet states. 
\end{abstract}    


\maketitle

% 1 -----------------------------------------------------------
\section{Introduction}
\label{sec:introduction}

Quantum one-dimensional (1D) spin systems have been 
studied in various aspects, especially with interest on their 
strong quantum fluctuations due to the low dimensionality. 
There appear a variety of quantum disordered ground states where 
the continuous spin rotation symmetry is not broken 
and the lowest spin excitation has a finite gap (spin-gap). 
These quantum disordered states have no analogues 
in classical spin systems. 
Typical examples are 
a Haldane state in a spin-1 chain~\cite{Haldane}, 
a dimer state in a spin-$\frac{1}{2}$ chain 
with bond alternation~\cite{b_alternation}, 
and a spin-gap state in a spin-$\frac{1}{2}$ ladder~\cite{ladder}. 

In extensive research for various 1D spin systems, 
spin chains with side chains 
have not attracted enough attention 
in spite of its potentially rich physics. 
Since a side chain is of finite length, 
it may enhance quantum fluctuation in the system. 
Actually,  the 1D Kondo necklace model, 
which has been extensively studied as a simplified version  of 
1D Kondo lattice model~\cite{Doniach,k_necklace}, 
can be regarded as a spin chain with side chains. 
In this model, the main chain is a spin-$\frac{1}{2}$ chain 
and each side chain consists of a single spin with 
magnitude $\frac{1}{2}$. 
The ground state of this model is known to be 
in the Kondo singlet phase with spin-gap~\cite{k_necklace}, 
while the spin-$\frac{1}{2}$ chain without  side chains is critical. This means that  the quasi-long range order in the main chain 
is destroyed by the quantum fluctuation in the side chains. 
Also, if the side chains bring geometrical frustration 
into the system, quantum fluctuation 
is expected to be further enhanced. 
Thus, it is an interesting subject how quantum fluctuation 
manifests itself and what kind of ground state appears 
in various types of spin chains with side chains. 

\begin{figure}[b]
\begin{center}\leavevmode
\includegraphics[width=0.6\linewidth]{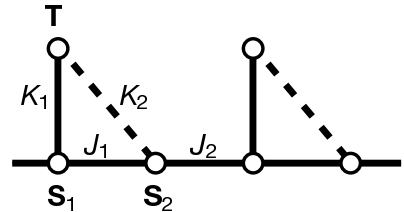}
\end{center}
\caption{A quantum spin chain with side chains; 
two unit cells are presented. 
$\v{S}_1$, $\v{S}_2$, and $\v{T}$ are 
spins whose magnitudes satisfy Eq.~(\ref{restriction}). 
The case that $S_1=1$, 
$S_2=\frac{1}{2}$, and $T=\frac{1}{2}$ is 
studied particularly in detail.} 
\label{lattice}
\end{figure}

In this paper, we investigate the natures of 
quantum disordered ground states of 
one of the simplest models with frustrated side chains. 
The main chain of the model consists of two species of spins 
in alternating order, and each side chain consists of a single spin 
which are alternately attached to the main chain, 
as will be shown in Fig.~\ref{lattice} of the next section. 
This model incorporates the effects of mixture of 
different spins, bond alternation, and frustration 
in spite of its simplicity. 
In particular, the frustration comes from triangles, 
each consisting of antiferromagnetically interacting 
three spins. 
Although quantum spin systems with similar geometry 
have been investigated by several authors~\cite{chain_triangle}, 
our model is physically different from them. 
If the side chains of our model are removed, 
the main chain is in a ferrimagnetic ground state.  
The side chain spins introduce frustration to the system, 
and destroy the ferrimagnetic order 
leading to quantum disordered states. 

%-->-->-->-->-->-->-->-->-->-->-->-->-->-->-->-->
%Fig. 1
%--<--<--<--<--<--<--<--<--<--<--<--<--<--<--<--<

The disordered ground states of the present model cannot 
be understood in the conventional valence bond solid 
(VBS) picture~\cite{AKLT}, which successfully explains 
the disordered ground states of many spin models 
with local frustration. 
Instead, we will explain the present ground states 
in the concept of the singlet cluster solid (SCS) picture. 
A SCS state is a direct product of local singlet states, 
or singlet clusters. 
Each singlet cluster consists of more than two singlet dimers, 
and the dimers are resonating locally within the cluster. 
The SCS states are realized as a result of the interplay of 
quantum fluctuation and local frustration, 
as will be explained in detail. 

It is desirable that the SCS state manifested in this paper 
is experimentally inspected in materials. 
However, a material precisely described by the present model 
is not found so far within our knowledge. 
Despite the lack of materials, 
it is worth clarifying the concept of the SCS states and 
verifying the existence of them in a concrete model. 
Further, considering the rich variety of magnetic materials synthesized by the modern chemical technology \cite{Miller}, 
desired materials are expected to be synthesized, 
since they are not necessarily complex in structure. 

The details of the model are explained in the next section. 
Various approaches are employed to clarify the ground state 
phases of this model: 
In section \ref{sec:nlsm}, a nonlinear $\sigma$ model (NLSM) method is proposed to grasp qualitative feature of the phase 
diagram. 
Since a similar NLSM method has been developed for 
mixed spin chains without side chains so far, 
we extend it to the present side-chain case. 

We also employ other approaches limiting ourselves to 
the simplest case of spin magnitudes 1 and $\frac{1}{2}$. 
In section \ref{sec:num_diagonal}, the numerical diagonalization 
for finite systems is carried out to obtain 
the quantitatively reliable phase diagram, 
which confirms the qualitative correctness of the NLSM method. 
Some limiting cases are exactly treated to draw physical picture 
for each phase in section \ref{sec:limit}. 
After these considerations, we arrive at the SCS pictures 
to explain the ground states in section \ref{sec:scs}. 
The SCS pictures are also supported by variational calculations 
in section \ref{sec:variational}. 
The last section is devoted to summary and discussion. 

% 2 --- Hamiltonian --------------------------
\section{Model Hamiltonian and its Classical Version}
\label{sec:hamiltonian}

We study an isotropic quantum spin chain 
with  alternately arrayed side chains 
as is illustrated in Fig.~\ref{lattice}. 
In the $p$th unit cell, $\v{S}_1(p)$ and 
$\v{S}_2(p)$ are spin operators on the main chain, 
and $\v{T}(p)$ is a spin operator 
on the side chain. 
The quantum numbers of spin magnitudes of these spin operators 
are expressed as $S_1$, $S_2$, and $T$, respectively. 
Exchange parameters are represented as $J_1$, $J_2$, $K_1$ 
and $K_2$, and assumed to be all positive. 
Then the Hamiltonian is written as 
%>>>>>>>>>>>>>>>>>>>>>>>>>>>>>>>>>>>>>>>>>>>
\begin{align} %(1)
H &= \sum_{p=1}^{N} \left\{
J_1 \, \v{S}_1(p) \cdot \v{S}_2(p) + 
J_2 \, \v{S}_2(p) \cdot \v{S}_1(p+1) 
\right. 
\nonumber \\
& \quad + \left. K_1 \, \v{S}_1(p) \cdot \v{T}(p) 
+ K_2 \, \v{S}_2(p) \cdot \v{T}(p) \right\} . 
\label{Hamiltonian}
\end{align}
%<<<<<<<<<<<<<<<<<<<<<<<<<<<<<<<<<<<<<<<<<<<<
The spacing between nearest spins is $a$, and the length of 
a unit cell is $2a$. 
The Hamiltonian is characterized by three independent dimensionless 
parameters: 
%>>>>>>>>>>>>>>>>>>>>>>>>>>>>>>>>>>>>>>>>>>>
\begin{align} %(2)
j = \frac{J_2}{J_1}, \quad 
k = \frac{T K_1}{S_2 J_1}, \quad 
r = \frac{S_2 K_2}{S_1 K_1}. 
\label{parameter}
\end{align}
%<<<<<<<<<<<<<<<<<<<<<<<<<<<<<<<<<<<<<<<<<<<<
Here $j$ measures the strength of the bond alternation 
in the main chain, $k$ measures the strength of interaction 
between a main-chain spin and a side-chain spin, and 
$r$ measures the strength of frustration~\cite{simple_chain}. 

In the present paper, we assume the following restriction on spin magnitudes: 
%>>>>>>>>>>>>>>>>>>>>>>>>>>>>>>>>>>>>>>>>>>>
\begin{align} %(3)
S_1 - S_2 - T = 0 .  
\label{restriction}
\end{align}
%<<<<<<<<<<<<<<<<<<<<<<<<<<<<<<<<<<<<<<<<<<<<
This is the condition that the corresponding classical spin chain 
can have a ground state with  
no total magnetization (i.~e. no ferrimagnetism), 
when $K_2$ is not large. 
The restriction (\ref{restriction}) serves to simplify 
the Berry phase term in the continuum limit. 

Expectation values of the spin operators for a spin coherent state 
are written as
%>>>>>>>>>>>>>>>>>>>>>>>>>>>>>>>>>>>>>>>>>>>
\begin{align} %(4)
\langle \v{S}_1(p) \rangle &= S_1 \v{M}_1(p) , 
\nonumber \\
\langle \v{S}_2(p) \rangle &= - S_2 \v{M}_2(p) , 
\nonumber \\
\langle \v{T}(p) \rangle &= - T \v{M}_{\bot}(p) , 
\label{coherent}
\end{align}
%<<<<<<<<<<<<<<<<<<<<<<<<<<<<<<<<<<<<<<<<<<<<
where $\v{M}_1(p)$, $\v{M}_2(p)$, and 
$\v{M}_{\bot}(p)$ are unit vectors. 
Replacing the spin operators in Eq.~(\ref{Hamiltonian}) 
by them, we have the classical version of the Hamiltonian: 
%>>>>>>>>>>>>>>>>>>>>>>>>>>>>>>>>>>>>>>>>>>>
\begin{align} %(5)
H_c &= \frac{1}{2}\tilde{J}_1 \sum_{p=1}^{N} 
\Bigl\{
j \, [ \v{M}_2(p) - \v{M}_1(p+1) ]^2 
\Bigr. 
\nonumber \\
& \ \ + \left(1-\frac{kr}{1-r} \right) \, 
[ \v{M}_1(p) - \v{M}_2(p) ]^2 
\nonumber \\
& \ \ + \Bigl. \frac{k}{1-r} \, 
[ \v{M}_1(p) - r\v{M}_2(p) 
- (1-r)\v{M}_{\bot}(p) ]^2 
\Bigr\} , 
\nonumber \\
\label{c_Hamiltonian}
\end{align}
%<<<<<<<<<<<<<<<<<<<<<<<<<<<<<<<<<<<<<<<<<<<<
where $\tilde{J}_1 = J_1 S_1 S_2$ and a constant term 
is omitted from $H_c$. 
The classical antiferromagnetic configuration, 
%>>>>>>>>>>>>>>>>>>>>>>>>>>>>>>>>>>>>>>>>>>>
\begin{align} %(6)
\v{M}_1(p) = \v{M}_2(p') = \v{M}_{\bot}(p'') 
\label{c_solution}
\end{align}
%<<<<<<<<<<<<<<<<<<<<<<<<<<<<<<<<<<<<<<<<<<<<
for all $p$, $p'$, and $p''$, is the ground-state solution  
if the pre-factors of the squares 
in Eq.~(\ref{c_Hamiltonian}) are all positive. 
This gives the condition for the classical stability of the antiferromagnetism as 
%>>>>>>>>>>>>>>>>>>>>>>>>>>>>>>>>>>>>>>>>>>>
\begin{align} %(7)
0 < r < 1 \, , \quad 0 < k < \frac{1}{r} -1 . 
\label{stability}
\end{align}
%<<<<<<<<<<<<<<<<<<<<<<<<<<<<<<<<<<<<<<<<<<<<
In the following arguments, 
we will concentrate on this region. 

% 3 --- NLSM ------------------------
\section{Nonlinear $\sigma$ Model for the Spin Chain}
\label{sec:nlsm}

Using the spin coherent representation, 
the partition function of Hamiltonian~(\ref{Hamiltonian}) 
is written in a path-integral form as 
%>>>>>>>>>>>>>>>>>>>>>>>>>>>>>>>>>>>>>>>>>>>
\begin{align} %(8)
Z = \int \prod_{j} D[\v{M}_j] \, 
\delta(\v{M}_j^2 - 1) e^{-A} 
\label{partition}
\end{align}
%<<<<<<<<<<<<<<<<<<<<<<<<<<<<<<<<<<<<<<<<<<<<
with $j = 1, 2$, and ${\bot}$. 
The action $A$ is written as 
%>>>>>>>>>>>>>>>>>>>>>>>>>>>>>>>>>>>>>>>>>>>
\begin{align} %(9)
A &= - iA_\mathrm{B} + A_\mathrm{H} , \nonumber \\ 
A_B &=  \sum_p \{ S_1 w[\v{M}_1(p)] 
- S_2 w[\v{M}_2(p)] 
- \ T w[\v{M}_{\bot}(p)] \} , \nonumber \\ 
A_H &= \int_0^{\beta} d\tau H_c , 
\label{action_coherent}
\end{align}
%<<<<<<<<<<<<<<<<<<<<<<<<<<<<<<<<<<<<<<<<<<<<
where $\beta$ is the inverse of temperature, and 
$w[\v{M}_i(p)]$ is the solid angle which 
$\v{M}_i(p)$ forms in period $\beta$. 
The term $- iA_B$ in action $A$ is the Berry phase term. 

We introduce slow variable $\v{m}(p)$ 
for each unit cell, and fluctuation variables 
$\v{L}_1(p)$, $\v{L}_2(p)$, and 
$\v{L}_{\bot}(p)$ 
for each spin in a unit cell. 
Then the original variables are transformed as follows: 
%>>>>>>>>>>>>>>>>>>>>>>>>>>>>>>>>>>>>>>>>>>>
\begin{align} %(10)
\v{M}_1(p) &= \v{m}(p) + a \v{L}_1(p) , 
\nonumber \\ 
\v{M}_2(p) &= \frac{1}{2} \v{m}(p) 
+ \frac{1}{2} \v{m}(p+1) + a \v{L}_2(p) , 
\nonumber \\ 
\v{M}_{\bot}(p) &=  \v{m}(p) 
+ a \v{L}_{\bot}(p) . 
\label{transformation}
\end{align}
%<<<<<<<<<<<<<<<<<<<<<<<<<<<<<<<<<<<<<<<<<<<<
This transformation is found by observing and extending 
the transformation for a simple chain without 
side chain~\cite{Takano,degree_f}. 
Since the left hand sides in (\ref{transformation}) are 
unit vectors, we have the following constraints for 
the new variables: 
%>>>>>>>>>>>>>>>>>>>>>>>>>>>>>>>>>>>>>>>>>>>
\begin{align} %(11)
\v{m}^2 = 1 \, , \quad 
\v{m} \cdot \v{L}_1 = 
\v{m} \cdot \v{L}_2 = 
\v{m} \cdot \v{L}_{\bot} = 0 . 
\label{constraint}
\end{align}
%<<<<<<<<<<<<<<<<<<<<<<<<<<<<<<<<<<<<<<<<<<<<
The fluctuation variables depends on one another, 
and one of them, e.~g. $\v{L}_{\bot}$ can be set equal to 0. 
Hence the number of independent variables are conserved 
in the transformation. 
Defining new fluctuation variables, 
%>>>>>>>>>>>>>>>>>>>>>>>>>>>>>>>>>>>>>>>>>>>
\begin{align} %(12)
\label{fluctuation_variable}
\v{R} &= \v{L}_2 - \v{L}_1 , 
\nonumber \\
\v{Q} &= (1-r)\v{L}_{\bot}  - \v{L}_1 
+ r \v{L}_2 , 
\end{align}
%<<<<<<<<<<<<<<<<<<<<<<<<<<<<<<<<<<<<<<<<<<<<
and taking the continuum limit, we have 
%>>>>>>>>>>>>>>>>>>>>>>>>>>>>>>>>>>>>>>>>>>>
\begin{align} %(13)
\label{action-continuum}
 A &= \int \!\! d\tau dx 
\Biggr\{ 
\frac{i}{2} S_2 
\frac{\partial \v{m} }{\partial x} \cdot 
\left( \v{m} \times 
\frac{\partial \v{m} }{\partial \tau}\right) 
\Biggl. 
\nonumber \\ 
& \ \ + \frac{a}{4} \tilde{J}_1 (c_+ - kr^2)  
\left( \frac{\partial \v{m} }{\partial x} \right)^2 
\nonumber \\ 
& \ \ +  \Biggl. \frac{a}{4} \tilde{J}_1 
[ k (\v{Q}^2 + 2\v{f} \cdot \v{Q})  
+ c_+ (\v{R}^2 + 2\v{g} \cdot \v{R}) ] 
\Biggr\} 
\end{align}
%<<<<<<<<<<<<<<<<<<<<<<<<<<<<<<<<<<<<<<<<<<<<
with $c_{\pm} = 1 \pm j -kr/(1-r)$. 
Here vectors $\v{f}$ and $\v{g}$ are given as 
%>>>>>>>>>>>>>>>>>>>>>>>>>>>>>>>>>>>>>>>>>>>
\begin{align} %(14)
\label{f_and_g}
 \v{f} &= r \frac{\partial \v{m} }{\partial x} 
+ i \frac{S_1 - S_2}{a \tilde{J}_1 k (1-r)} 
\, \v{m} \times \frac{\partial \v{m} }{\partial \tau} , 
 \nonumber \\ 
 \v{g} &= 
\frac{c_-}{c_+} \frac{\partial \v{m} }{\partial x} 
- i \frac{rS_1 - S_2}{c_+ a \tilde{J}_1 (1-r)} 
\, \v{m} \times \frac{\partial \v{m} }{\partial \tau} . 
\end{align}
%<<<<<<<<<<<<<<<<<<<<<<<<<<<<<<<<<<<<<<<<<<<<

Integrating the partition function with respect to 
$\v{R}$ and $\v{Q}$, 
we have the following NLSM action: 
%>>>>>>>>>>>>>>>>>>>>>>>>>>>>>>>>>>>>>>>>>>>
\begin{align} %(15)
\label{action-NLSM}
 A_\mathrm{eff} &={} \int \!\! d\tau \!\! \int \!\! dx 
\Biggl\{ 
i \frac{\theta}{4\pi} 
\v{m} \cdot 
\left( \frac{\partial \v{m} }{\partial \tau}
\times \frac{\partial \v{m} }{\partial x} \right) 
\Biggr. 
\nonumber \\ 
& \ \ + \frac{c_+ (S_1 -  S_2)^2 + k (rS_1 -  S_2)^2}
{4a\tilde{J}_1(1-r)^2 k c_+} 
\left( \frac{\partial \v{m} }{\partial \tau} \right)^2 
\nonumber \\ 
& \ \ + \Biggl. \frac{a}{4} \tilde{J}_1 c_+ 
\left( 1 - \frac{c_-^2}{c_+^2} \right) 
\left( \frac{\partial \v{m} }{\partial x} \right)^2 
\Biggr\} .  
\end{align}
%<<<<<<<<<<<<<<<<<<<<<<<<<<<<<<<<<<<<<<<<<<<<
The first term is the topological term 
and $\theta$ is the topological angle given by 
%>>>>>>>>>>>>>>>>>>>>>>>>>>>>>>>>>>>>>>>>>>>
\begin{align} %(16)
\theta = \frac{4\pi j (S_2 - r S_1)}{(1-r) (1+j)-kr} . 
\label{t_angle}
\end{align}
%<<<<<<<<<<<<<<<<<<<<<<<<<<<<<<<<<<<<<<<<<<<<

In the absence of frustration ($r=0$), the topological angle 
reduces to $\theta = 4\pi S_2 J_2 / (J_1 + J_2)$. 
This is the same expression as that for a simple spin chain 
of magnitude $S_2$ with 
bond alternation~\cite{Affleck,Takano}. 
This can be interpreted as follows: 
a spin $\v{S}_1$ is combined 
with the adjacent spin $\v{T}$ on the side chain, 
and a two-spin cluster with total spin magnitude 
$S_1 - T$ $(= S_2)$ is formed. 
However, the coefficients of 
$(\partial \v{m}/ \partial \tau)^2$ and 
$(\partial \v{m}/ \partial x)^2$ in 
Eq.~(\ref{action-NLSM}) do not reduce to  
those for the simple spin chain, 
even if $k$ is very large. 
This reflects that the quantum fluctuation of spins on side chains 
still survive for large $k$. 

The topological term of Eq.~(\ref{action-NLSM}) 
determines whether or not the system has a spin-gap 
in the same manner as Haldane's 
argument~\cite{Haldane}. 
That is, the system does not have a spin-gap if and only if 
the topological angle $\theta$ is just $\pi$ (mod $2\pi$). 
This condition 
is written as  
%>>>>>>>>>>>>>>>>>>>>>>>>>>>>>>>>>>>>>>>>>>>
\begin{align} %(17)
\frac{2j (S_2 - r S_1)}{(1-r) (1+j)-kr} = h, 
\label{gapless_condition}
\end{align}
%<<<<<<<<<<<<<<<<<<<<<<<<<<<<<<<<<<<<<<<<<<<<
where $h$ is any half odd integer. 
This gapless condition determines phase boundaries 
between gapful disordered phases 
in the parameter space. 
We notice that, for each value of $h$, 
boundaries for all values of $r$ pass through 
the common point 
%>>>>>>>>>>>>>>>>>>>>>>>>>>>>>>>>>>>>>>>>>>>
\begin{align} %(18)
(k, j) = \left( \frac{2(S_1 - S_2)}{2S_2 -h},  
\frac{h}{2S_2 -h} \right) . 
\label{common_point}
\end{align}
%<<<<<<<<<<<<<<<<<<<<<<<<<<<<<<<<<<<<<<<<<<<<

%-->-->-->-->-->-->-->-->-->-->-->-->-->-->-->-->
%Fig. 2
\begin{figure}[t]
\begin{center}\leavevmode
\includegraphics[width=0.85\linewidth]{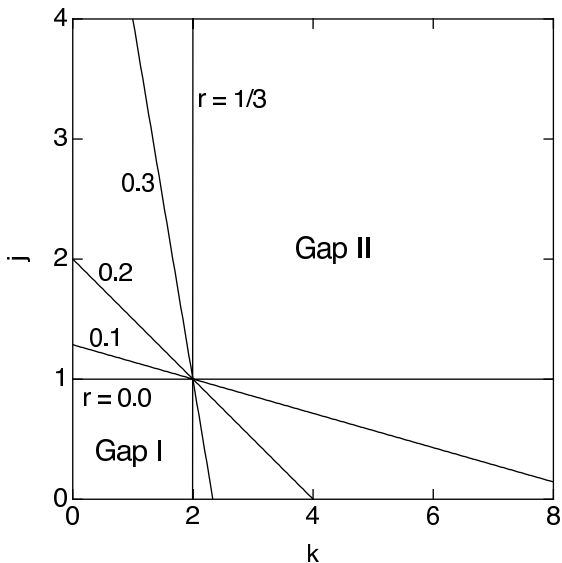}
\end{center}
\caption{Phase boundaries for several values of $r$ 
in the $k$-$j$ plane by the NLSM method 
for $S_1=1$, $S_2=\frac{1}{2}$, and $T=\frac{1}{2}$.
For each value of $r$ (= $K_2 / 2K_1$) 
the region of $0 < k < \frac{1}{r} -1$ is meaningful 
in the NLSM method. 
} 
\label{phase_nlsm}
\end{figure}
%--<--<--<--<--<--<--<--<--<--<--<--<--<--<--<--<

In the case of $S_1=1$, $S_2=\frac{1}{2}$ 
and $T=\frac{1}{2}$, only a permitted value of $h$ in 
Eq.~(\ref{gapless_condition}) is $\frac{1}{2}$ 
for $k>0$ and $j>0$. 
Then the phase boundaries for several values of $r$ 
are solid lines in Fig.~\ref{phase_nlsm}. 
Owing to the definitions of the parameters, 
they are straight lines in the present approximation. 
The regions of the both sides of each boundary are 
gapful disordered phases. 
We call them Gap~I phase and Gap~II phase as 
noted in the figure. 
For $r=0$ the phase boundary is horizontal, 
since the topological angle is independent of $k$ 
as mentioned below Eq.~(\ref{t_angle}). 
This is understandable by considering $\v{S}_1$ 
as a composite of two $\frac{1}{2}$ spins. 
In fact, when $r=0$, $\v{T}$ and one of 
the $\frac{1}{2}$ spins of $\v{S}_1$ necessarily 
forms a valence bond irrespective of the value of $k$ 
in the ground state. 
As $r$ increases from 0, the slope of the boundary 
becomes negatively large. 
The reason will be argued later. 

%-->-->-->-->-->-->-->-->-->-->-->-->-->-->-->-->
%Fig. 3
\begin{figure}[b]
\begin{center}\leavevmode
\includegraphics[width=1.00\linewidth]{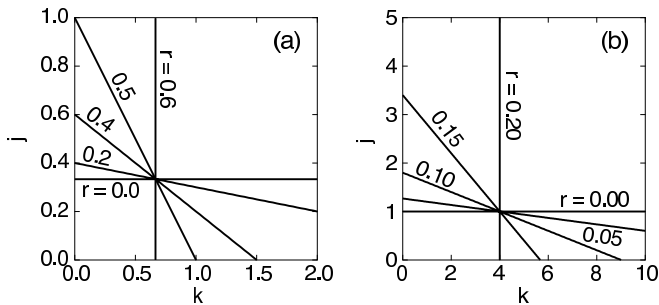}
\end{center}
\caption{Phase boundaries in the $k$-$j$ plane 
by the NLSM method 
in the cases of (a) $S_1=\frac{3}{2}$, 
$S_2=1$, and $T=\frac{1}{2}$, and (b) 
in the case of $S_1=\frac{3}{2}$, 
$S_2=\frac{1}{2}$, and $T=1$. 
For each value of $r$ 
the region of $0 < k < \frac{1}{r} -1$ is meaningful. 
} 
\label{other_phases}
\end{figure}
%--<--<--<--<--<--<--<--<--<--<--<--<--<--<--<--<

We also show phase diagrams in 
other cases in Fig.~\ref{other_phases}; 
the phase boundaries in (a) are for 
$S_1=\frac{3}{2}$, $S_2=1$, and $T=\frac{1}{2}$, 
and those in (b) are for 
$S_1=\frac{3}{2}$, $S_2=\frac{1}{2}$, and $T=1$. 
Although the equation (\ref{gapless_condition}) determining 
phase boundaries are quite general, we mainly 
examine the case of $S_1=1$, 
$S_2=\frac{1}{2}$, and $T=\frac{1}{2}$. 
This case is expected to include the essence of the present type of 
spin chains with side chains.

% 4 --- Numerical Diagonalization --------------------------
\section{Numerical Diagonalization}
\label{sec:num_diagonal}

Hamiltonian (\ref{Hamiltonian}) can be numerically 
diagonalized for small size systems. 
The numerical calculation is effective not only to 
analyze the system itself in detail, but also 
to know the preciseness of the NLSM method 
by comparing the results. 
We performed numerical diagonalization in the case of 
$S_1=1$, $S_2=\frac{1}{2}$, and $T=\frac{1}{2}$ 
to obtain the phase diagram for the ground state. 

%-->-->-->-->-->-->-->-->-->-->-->-->-->-->-->-->
%Fig. 4
\begin{figure}[b]
\begin{center}\leavevmode
\includegraphics[width=0.85\linewidth]{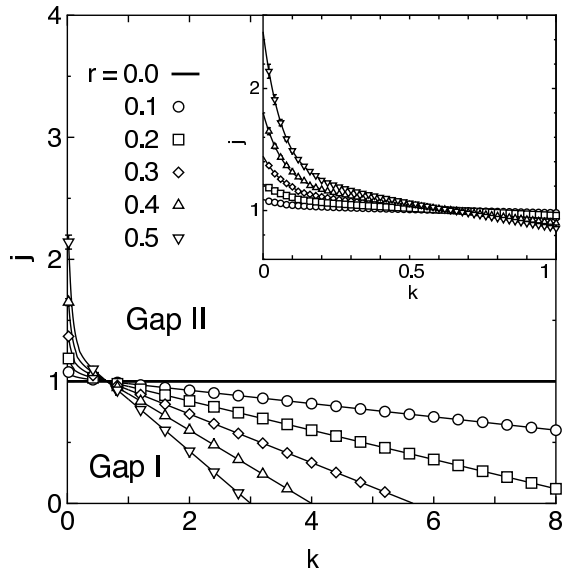}
\end{center}
\caption{Phase boundaries 
by numerical diagonalization for $S_1=1$, 
$S_2=\frac{1}{2}$, and $T=\frac{1}{2}$.
Each point of a boundary is determined by extrapolation 
for the total spin number $3N = $ 12, 18, and 24. 
Inset: Phase boundaries for small $k$. 
} 
\label{phase_num}
\end{figure}
%--<--<--<--<--<--<--<--<--<--<--<--<--<--<--<--<

The phase transition points are determined as follows. 
The phase transitions between different spin gap phases are 
expected to be the Gaussian transition. 
Hence we employ the method of twist boundary condition 
proposed by Kitazawa~\cite{Kitazawa} and 
Kitazawa and Nomura~\cite{Kita_Nomu} to determine 
the phase boundary. 
As will be examined later, the ground state phases are 
described by different SCS configurations. 
Under the twisted boundary condition, the different singlet solid configurations have different time reversal parities 
depending on the even-odd parity of the number of valence bonds 
across the twisted boundary. 
Hence the energy levels of the ground state and 
the first excited state cross at the phase boundary 
without level repulsion. 
This ensures the precise evaluation of the phase boundary. 
The size extrapolation is based on the following 
formula for the finite size correction\cite{Kitazawa,Kita_Nomu}: 
%>>>>>>>>>>>>>>>>>>>>>>>>>>>>>>>>>>>>>>>>>>>
\begin{align} %(19)
j_\mathrm{c}(\infty) = j_\mathrm{c}(N) + \frac{c_1}{N^2} 
+ \frac{c_2}{N^4} , 
\label{extrapolation}
\end{align}
%<<<<<<<<<<<<<<<<<<<<<<<<<<<<<<<<<<<<<<<<<<<<
where $j_\mathrm{c}(N)$ is the finite size critical value of 
quantity $j$, 
and $c_1$ and $c_2$ are fitting parameters. 
We have carried out the extrapolation using numerical results 
for total spin number $3N = $ 12, 18, and 24. 

Resultant phase boundaries for several values of $r$ 
are plotted in Fig.~\ref{phase_num}. 
Comparing Figs.~\ref{phase_nlsm} and \ref{phase_num}, 
we find that the NLSM method gives qualitatively correct 
phase boundaries. 
In particular, for small $r$, or weak frustration, 
the NLSM method provides a quantitatively fair approximation. 
With the increase of $r$, the Gap~I phase extends to 
the region $j > 1$ for small $k$, 
and it is suppressed to the region $j <1$ for large $k$.  
Although this feature qualitatively coincides with that of 
the NLSM results, quantitative coincidence becomes worse 
with the increase of $r$. 
This is natural because the present NLSM method starts 
from a classical antiferromagnetic solution in the absence of 
frustration. 

The possibility of the first order transition between different spin-gap phases has been pointed out in the frustrated ladder by Hakobyan and coworkers~\cite{Hakobyan} in the appropriate parameter regime. Considering the presence of frustration, this type of  transition cannot be ruled out in the present model. However, we did not find the numerical evidence for the first order transition within the parameter regime discussed in this paper.

% 6 --- Limiting Cases --------------------------
\section{Limiting Cases}
\label{sec:limit}

To further confirm the numerical phase diagram for 
$S_1=1$, $S_2=\frac{1}{2}$, and $T=\frac{1}{2}$, 
we consider the effective theory in the limiting cases of 
$j\rightarrow 0$ and $j \rightarrow \infty$. 

%=== subsection A
\subsection{Strong $J_1$ limit ($j \rightarrow 0$)}
\label{subsec:zero_limit}

\begin{figure}[t]
\begin{center}\leavevmode
\includegraphics[width=0.8\linewidth]{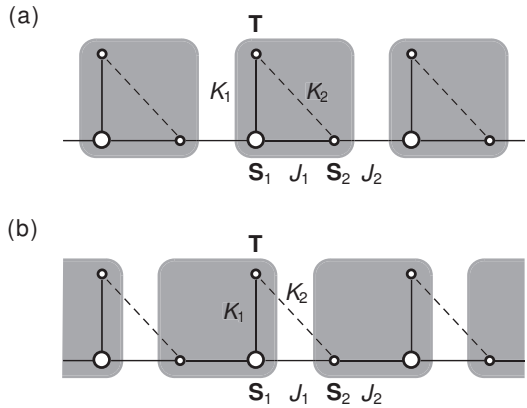}
\end{center}
\caption{(a) 3-spin units for $j\rightarrow 0$, and 
(b) those for $j \rightarrow \infty$ and $r \rightarrow 0$ 
in the case of $S_1=1$, $S_2=\frac{1}{2}$, 
and $T=\frac{1}{2}$.  
Three spins (circles) in a shadowed region form a 3-spin unit. 
} 
\label{3spin}
\end{figure}

In the limit of $j \rightarrow 0$, the system can be regarded 
as a one-dimensional array of weakly coupled 3-spin units 
as shown in Fig.~\ref{3spin}(a). 
One of the 3-spin units is described by the Hamiltonian 
%>>>>>>>>>>>>>>>>>>>>>>>>>>>>>>>>>>>>>>>>>>>
\begin{align} %(20)
\label{model2}
H_3 = J_1\v{S}_1\cdot\v{S}_2 
+ K_1\v{S}_1\cdot\v{T} 
+ K_2\v{S}_2\cdot\v{T}, 
\end{align}
%<<<<<<<<<<<<<<<<<<<<<<<<<<<<<<<<<<<<<<<<<<<<
where we have dropped the common index $p$ representing 
the $p$th unit cell for simplicity. 
The N\'eel basis is represented as $\ket{S^z_1, S^z_2, T^z}$, 
where $S^z_1$ takes $\Uparrow$, 0 or $\Downarrow$, 
and $S^z_2$ and $T$ take $\uparrow$ or $\downarrow$. 
By introducing the composed spin 
$\tilde{\v{S}}$ $\equiv$ 
$\v{S}_{1} + \v{S}_{2} +  \v{T}$, 
we have another set of basis vectors 
$\parallel\! \tilde{S} , \tilde{S}^z,\alpha \rangle\!\rangle$, 
where $\tilde{S}$ and $\tilde{S}^z$ are quantum numbers of 
the magnitude and the $z$-component of $\tilde{\v{S}}$ 
respectively, and $\alpha$ discriminates multiple states with 
the same $\tilde{S}$ and $\tilde{S}^z$, if necessary. 
We seek the ground state of the 3-spin unit 
with $\tilde{S}=0$ or 1, 
since all the exchange interactions are antiferromagnetic. 

%-->-->-->-->-->-->-->-->-->-->-->-->-->-->-->-->
%Fig. 5
%--<--<--<--<--<--<--<--<--<--<--<--<--<--<--<--<

(i) For $\tilde{S}=0$, we have $\tilde{S}^z=0$. 
Then the one-dimensional subspace consists of a single state,
%>>>>>>>>>>>>>>>>>>>>>>>>>>>>>>>>>>>>>>>>>>>
\begin{align} %(21)
\label{sing_wf}
\kket{0,0} &= \frac{1}{\sqrt{6}}
( \, \ket{0\downarrow\uparrow}
+\ket{0\uparrow\downarrow}
\nonumber\\ & \qquad 
-\sqrt{2}\ket{\Uparrow\downarrow\downarrow}
-\sqrt{2}\ket{\Downarrow\uparrow\uparrow}) . 
\end{align}
%<<<<<<<<<<<<<<<<<<<<<<<<<<<<<<<<<<<<<<<<<<<<
This is a singlet eigenstate of $H_3$ 
belonging to the eigenvalue 
%>>>>>>>>>>>>>>>>>>>>>>>>>>>>>>>>>>>>>>>>>>>
\begin{align} %(22)
E_0 = -J_1-K_1+\frac{K_2}{4} . 
\label{sing_ene}
\end{align}
%<<<<<<<<<<<<<<<<<<<<<<<<<<<<<<<<<<<<<<<<<<<<

(ii) For $\tilde{S}=1$, we have 
$\tilde{S}^z$ = 1, 0, or $-1$. 
In this three-dimensional subspace, 
it is sufficient to inspect the case of $\tilde{S}^z=1$ 
owing to the spherical symmetry of the Hamiltonian $H_3$. 
We choose the orthonormal basis of the subspace 
$\tilde{S}=\tilde{S}^z=1$ as 
%>>>>>>>>>>>>>>>>>>>>>>>>>>>>>>>>>>>>>>>>>>>
\begin{align} %(23)
\kket{1,1,1}&=\frac{1}{2}( \, \ket{\Uparrow\downarrow\uparrow}+\ket{\Uparrow\uparrow\downarrow}-\sqrt{2}\ket{0\uparrow\uparrow} \, ) , 
\nonumber \\
\kket{1,1,2}&=\frac{1}{\sqrt{2}}( \, \ket{\Uparrow\downarrow\uparrow}-\ket{\Uparrow\uparrow\downarrow} \, ).
\end{align}
%<<<<<<<<<<<<<<<<<<<<<<<<<<<<<<<<<<<<<<<<<<<<
Operating $H_3$ on these bases, we have an eigenvalue equation. 
Then the lowest eigenvalue $E_1$ in this subspace is determined 
as the smaller solution of the characteristic equation: 
%>>>>>>>>>>>>>>>>>>>>>>>>>>>>>>>>>>>>>>>>>>>
\begin{align} %(24)
\left|
\begin{array}{ll}
-\dfrac{J_1}{2}-\dfrac{K_1}{2}+\dfrac{K_2}{4}-E_1 & \dfrac{1}{\sqrt{2}}(-J_1+K_1)\\
\dfrac{1}{\sqrt{2}}(-J_1+K_1) & -\dfrac{3K_2}{4}-E_1\\
\end{array}
\right|=0 . 
\label{tri_ene}
\end{align}
%<<<<<<<<<<<<<<<<<<<<<<<<<<<<<<<<<<<<<<<<<<<<

Introducing a normalized energy difference as 
$\epsilon = (E_1-E_0)/J_1$, 
Eq.~(\ref{tri_ene}) with Eq.~(\ref{sing_ene}) reduces to 
%>>>>>>>>>>>>>>>>>>>>>>>>>>>>>>>>>>>>>>>>>>>
\begin{align} %(25)
\label{ene_diff}
2 \epsilon^2-[3(1+k)-4rk]\epsilon
+2k[2-\left(1+k\right)r]=0 . 
\end{align}
%<<<<<<<<<<<<<<<<<<<<<<<<<<<<<<<<<<<<<<<<<<<<
If the smaller solution for $\epsilon$ is negative, 
the ground state is a triplet ($\tilde{S}=1$) state; 
otherwise it is a singlet ($\tilde{S}=0$) state. 
Using (\ref{ene_diff}), the condition 
for the triplet ground state becomes
%>>>>>>>>>>>>>>>>>>>>>>>>>>>>>>>>>>>>>>>>>>>
\begin{align} %(26)
k > k_\mathrm{c}\equiv\frac{2}{r}-1 \quad (\tilde{S}=1) . 
\end{align}
%<<<<<<<<<<<<<<<<<<<<<<<<<<<<<<<<<<<<<<<<<<<<
We notice that $k_\mathrm{c}>1$ in the region of 
$0<r<1$, which we have concentrated on in this paper. 
In the triplet ground state, 
$\v{S}_2$ tends to orient 
the opposite direction to $\v{S}_1$ for $J_1 > K_1$, 
and $\v{T}$ does for $J_1 < K_1$. 
The composed spin $\tilde{\v{S}}$ 
always orients to the same direction as $\v{S}_1$. 

The composed spin $\tilde{\v{S}}(p)$ 
at the $p$th unit cell interacts with 
adjacent $\tilde{\v{S}}(p+1)$ 
by an effective exchange interaction. 
We denote the effective exchange parameter 
by $J_\mathrm{eff}$. 
Since the interaction  
between $\v{S}_2(p)$ and $\v{S}_1(p+1)$ is 
antiferromagnetic ($J_2 > 0$), 
the correlation between $\v{S}_1(p)$ and 
$\v{S}_1(p+1)$ 
is antiferromagnetic for $K_1 > J_1$ $(k>1)$ 
and ferromagnetic for $K_1 < J_1$ $(k<1)$. 
Therefore $J_\mathrm{eff}$ has the same sign as 
$K_1 - J_1 = J_1(k-1)$, considering the signs of  
$\aver{\tilde{\v{S}}(p) \cdot \tilde{\v{S}}(p+1)}$ 
and $\aver{\v{S}_1(p) \cdot \v{S}_1(p+1)}$ are the same. 

For $k > k_\mathrm{c}$, we have $J_\mathrm{eff} > 0$, 
since $k_\mathrm{c} > 1$ for $0<r<1$. 
Hence the original spin chain is equivalent to 
a spin-1 antiferromagnetic Heisenberg chain consisting of 
effective spins, $\tilde{\v{S}}(p)$'s. 
The ground state of an uniform spin-1 chain is the Haldane 
state~\cite{Haldane}, 
which gives a spin-gap for excitation. 
In the Haldane state, there is strong correlation on 
each adjacent spin pair, as known from the VBS picture 
for effective spins, $\tilde{\v{S}}(p)$'s~\cite{AKLT}. 
In terms of the original spins, there is strong correlation  
between adjacent 3-spin units. 
For $k < k_c$, on the other hand, 
the ground state of each 3-spin unit is already 
a closed singlet state. 
Then the ground state of the total spin chain is approximately 
an array of such closed local singlets, and 
there is almost no correlation  between adjacent 3-spin units. 
Thus there is a Gaussian transition between 
the two characteristic ground states with spin-gap at $k=k_c$. 
The value of $k_c$ in this argument for $j \rightarrow 0$ 
agrees with the critical value by the numerical diagonalization, 
as is seen on the $j=0$ line of the phase diagram 
(Fig.~\ref{phase_num}).

%=== subsection B
\subsection{Strong $J_2$ limit ($j \rightarrow \infty$)}
\label{subsec:inf_limit}

In the limit of $j \rightarrow \infty$, spins 
$\v{S}_1(p+1)$ and $\v{S}_2(p)$ form 
an effective spin $\hat{\v{S}}(p)$ $\equiv$ 
$\v{S}_1(p+1)+\v{S}_2(p)$ with 
magnitude~$\frac{1}{2}$, 
and other interactions can be treated as perturbations. 
Then the effective Hamiltonian for $\hat{\v{S}}(p)$ 
and $\v{T}(p)$ is 
%>>>>>>>>>>>>>>>>>>>>>>>>>>>>>>>>>>>>>>>>>>>
\begin{align} %(27)
H_\mathrm{eff} 
&= \sum_{p=1}^{N} \left\{
-\frac{4}{9} J_1 \, \hat{\v{S}}(p) \cdot 
\hat{\v{S}}(p+1)
\right. 
\nonumber \\
& + \left. \frac{4}{3} K_1 \, 
\hat{\v{S}}(p) \cdot \v{T}(p+1) 
- \frac{1}{3} K_2  \, 
\hat{\v{S}}(p) \cdot \v{T}(p) \right\} . 
\label{heff} 
\end{align}
%<<<<<<<<<<<<<<<<<<<<<<<<<<<<<<<<<<<<<<<<<<<<
The ground state of this chain is still nontrivial. 
However, $K_2$ plays a secondary role 
in the weakly frustrated region, so that 
each $\v{T}$ antiferromagnetically interacts with 
the ferromagnetic chain consisting of $\hat{\v{S}}$'s.  
Therefore, it is plausible that the ground state is always 
nonmagnetic for small $r$.  
Numerical studies of the effective model (\ref{heff}) suggest 
no phase transition for $0 < r <1/3$ where no phase transition 
is predicted by the NLSM method for large $j$. 

In terms of the original Hamiltonian (\ref{Hamiltonian}), 
the ground state of $j \rightarrow \infty$ 
and $r \rightarrow 0$ 
is a direct product of local singlet states of 3-spin units.  
A 3-spin unit consists of $\v{S}_2(p)$, 
$\v{S}_1(p+1)$, and $\v{T}(p+1)$, 
as shown in Fig.~\ref{3spin}(b). 
The ground state of finite $k$ and $r$ ($j \rightarrow \infty$) 
is adiabatically connected to the limit without phase transition 
as long as $r$ is small.  The full phase diagram of the effective 
model (\ref{heff}) is investigated in a separate paper~\cite{delta}.

% 7 --- Singlet Cluster Solid Picture --------------------------
\section{Singlet Cluster Solid Picture}
\label{sec:scs}

%-->-->-->-->-->-->-->-->-->-->-->-->-->-->-->-->
%Fig. 6
\begin{figure}[t]
\begin{center}\leavevmode
\includegraphics[width=0.7\linewidth]{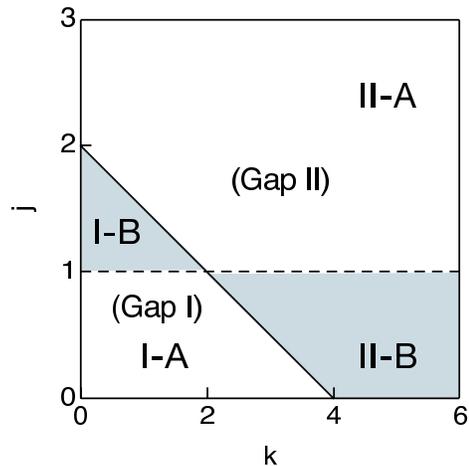}
\end{center}
\caption{
Four regions in a typical phase diagram for 
$S_1=1$, $S_2=\frac{1}{2}$, and $T=\frac{1}{2}$; 
the phase diagram for $r=0.2$ by the NLSM method is shown. 
The solid line is the phase boundary between Gap~I phase 
and Gap~II phase. 
In each phase, the dashed line of $j=1$ means a crossover 
between regions with different features. 
Gap~I phase consists of regions I-A and I-B, while 
Gap~II phase consists of regions II-A and II-B. 
They are explained by SCS pictures. 
} 
\label{regions}
\end{figure}
%--<--<--<--<--<--<--<--<--<--<--<--<--<--<--<--<

In this section, we propose the SCS picture 
to explain any ground state in the phase diagram 
for $S_1=1$, $S_2=\frac{1}{2}$, and $T=\frac{1}{2}$. 
This is a generalization of the VBS picture, 
and is based on expressing $\v{S}_1$ with magnitude 1 as 
%>>>>>>>>>>>>>>>>>>>>>>>>>>>>>>>>>>>>>>>>>>>
\begin{align} %(28)
\v{S}_1 = \v{S}_1^{(1)} + \v{S}_1^{(2)}, 
\end{align}
%<<<<<<<<<<<<<<<<<<<<<<<<<<<<<<<<<<<<<<<<<<<<
where $\v{S}_1^{(1)}$ and $\v{S}_1^{(2)}$ 
are spins with magnitude~$\frac{1}{2}$~\cite{symmetrize}. 

For convenience of explanation, we divide each phase into 
two regions by the line of $j=1$ 
as schematically shown in Fig.~\ref{regions}: 
Gap~I phase is divided into regions I-A and I-B, and 
Gap~II phase is divided into regions I-A and I-B. 

%=== subsection A
\subsection{VBS picture and its insufficiency}
\label{subsec:vbs}

%-->-->-->-->-->-->-->-->-->-->-->-->-->-->-->-->
%Fig. 7
\begin{figure}[t]
\begin{center}\leavevmode
\includegraphics[width=0.9\linewidth]{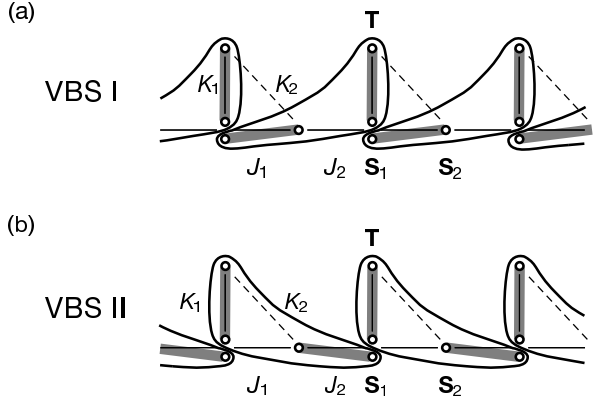}
\end{center}
\caption{VBS pictures in the limiting cases 
for $S_1=1$, $S_2=\frac{1}{2}$, and $T=\frac{1}{2}$: 
(a) VBS~I represents the ground state 
for $j\rightarrow 0$ and $k < k_c$, and 
(b) VBS~II does for $j\rightarrow \infty$. 
Small circles are $\frac{1}{2}$-spins, and 
bold gray lines are valence bonds. 
Spin $\v{S}_{1}$ is expressed by two 
$\frac{1}{2}$-spins as 
$\v{S}_1 = \v{S}_1^{(1)} + \v{S}_1^{(2)}$. 
Loops including two valence bonds are for correspondence to 
SCS pictures (see text). 
} 
\label{vbs}
\end{figure}
%--<--<--<--<--<--<--<--<--<--<--<--<--<--<--<--<

The ground states of the limiting cases in the preceding section 
are explained by VBS pictures. 
The VBS picture for $j\rightarrow 0$ and $k < k_c$ is 
VBS~I illustrated in Fig.~\ref{vbs}(a)~\cite{equiv_VBS_I}. 
The valence bonds on $J_1$-interactions, 
which we hereafter abbreviate as the $J_1$-valence-bonds, 
mainly contribute to the energy gain of the ground state of VBS~I. 
On the other hand, the VBS picture for $j \rightarrow \infty$ is 
VBS~II illustrated in Fig.~\ref{vbs}(b). 
The valence bonds on $J_2$-interactions, 
or the $J_2$-valence-bonds, 
mainly contribute to the energy gain of the ground state of VBS~II. 

The VBS pictures are not adequate for regimes 
away from the above limiting cases, 
although they are expected to be qualitatively valid 
for small-$j$ and large-$j$ regimes. 
As seen in Fig.~\ref{regions}, 
the energetic advantage of $J_1$-valence-bonds of 
VBS~I on line $j = 0$ ($k < k_c$) remains within region I-A 
because $J_1 > J_2$. 
However the advantage is lost in region I-B because 
$J_2 > J_1$. 
Similarly, the energetic advantage of $J_2$-valence-bonds of 
VBS~II in the limit of $j\rightarrow \infty$ remains 
within region II-A because $J_2 > J_1$. 
However the advantage is lost in region II-B because 
$J_1 > J_2$. 
Since the line of $j=1$ is not a phase boundary, 
we need a new picture to explain the whole Gap~I (II) phase 
which reduces to VBS~I (II) in the limit. 
The picture will be a SCS picture. 

%=== subsection B
\subsection{Concept of SCS picture}
\label{subsec:concept}

A general SCS picture is defined by a wave function 
of a tensor product form of local singlet states. 
We call this wave function the SCS state and 
each local singlet state a singlet cluster. 
It is typically written as 
%>>>>>>>>>>>>>>>>>>>>>>>>>>>>>>>>>>>>>>>>>>>
\begin{align} %(29)
\label{var_fun_gen}
\ket{\Psi} = \ket{\psi(1)} 
\otimes \ket{\psi(2)} \cdots 
\otimes \ket{\psi(M)} , 
\end{align}
%<<<<<<<<<<<<<<<<<<<<<<<<<<<<<<<<<<<<<<<<<<<
where $\ket{\psi(p)}$ $(p = 1, 2, \cdots , M)$ is 
a singlet cluster and 
$M$ is the total number of singlet clusters in the SCS state. 
Here we have considered that any spin with magnitude more than 1 
is resolved into a set of spins with magnitude~$\frac{1}{2}$. 
Then a singlet cluster is a singlet state of more than two spins 
with magnitude~$\frac{1}{2}$. 

A singlet cluster in a SCS state is represented as 
a superposition of products of valence bonds. 
Hence the valence bonds are resonating within the singlet cluster. 
A VBS is a special case of the SCS, where 
a singlet cluster is a single valence bond and 
no resonation occurs. 
A resonating valence bond (RVB) state is another special case, 
where the whole system is the singlet cluster 
and all valence bonds are resonating. 

Usually a SCS state is not the exact ground state 
for a given  spin Hamiltonian. 
However if the exact ground state is continuously 
modified into an appropriate SCS state, 
the SCS state describes the essence of the ground state. 
From this viewpoint, 
the SCS state is useful to characterize the phase 
which the ground state belongs to. 
In some quantum spin chains without side chain, 
various ground-state phases have been successfully described 
by corresponding SCS pictures~\cite{Takano2}. 
Furthermore, a SCS state can also quantitatively describe 
the true ground state, if the wave function of each singlet cluster 
is well localized. 
Among such systems, 
in the spin system on a diamond chain~\cite{Takano3}, 
we have the exact tetramer-dimer-state solution, 
which is a kind of SCS state.

%=== subsection C
\subsection{SCS picture for Gap~I phase}
\label{subsec:gap_I}

%-->-->-->-->-->-->-->-->-->-->-->-->-->-->-->-->
%Fig. 8
\begin{figure}[t]
\begin{center}\leavevmode
\includegraphics[width=0.9\linewidth]{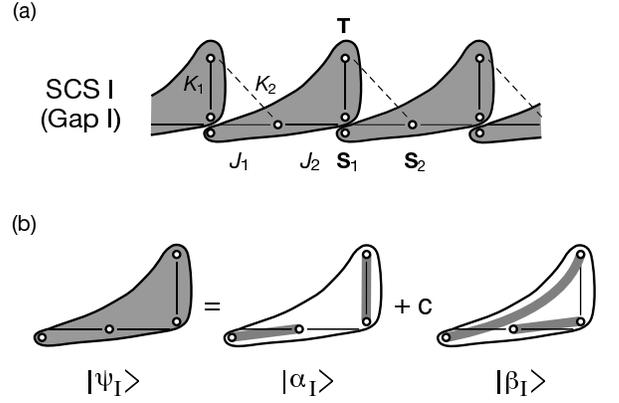}
\end{center}
\caption{(a) SCS~I, the SCS picture for Gap~I phase 
($S_1=1$, $S_2=\frac{1}{2}$, and $T=\frac{1}{2}$). 
SCS~I is a tensor product form of singlet clusters. 
(b) A singlet cluster $\ket{\psi_\mathrm{I}(c)}$ 
(the left hand side) in SCS~I. 
It is represented as a linear combination of 
two valence bond states, $\ket{\alpha_\mathrm{I}}$ and 
$\ket{\beta_\mathrm{I}}$ (the right hand side). 
$c$ is the coefficient of the linear combination. 
} 
\label{scs_I}
\end{figure}
%--<--<--<--<--<--<--<--<--<--<--<--<--<--<--<--<

We call the SCS picture for the Gap~I phase SCS~I. 
The SCS~I state is constructed by 
assuming  the following requirements: 

(i)  The ground state is continuously modified to 
the VBS~I state without global rearrangement of 
the valence bond configuration. 

(ii)  The ground state is invariant under the translation 
by a single unit cell. 

(iii) The ground state contains 
a substantial amount of component with 
$J_2$-valence-bonds in region I-B. 

Assumption (i) is necessary, since the Gap~I phase involves 
the limiting case of $j\rightarrow 0$ and $k < k_c$ where 
VBS~I picture holds, and 
there is no phase transition in Gap~I phase. 
As for assumption (ii), we confirmed that there is 
no indication of the translational symmetry breaking 
in numerical ground states. 
Assumption (iii) is required to explain region I-B. 

Under the above requirements, we take SCS~I wave function 
in the tensor product form:
%>>>>>>>>>>>>>>>>>>>>>>>>>>>>>>>>>>>>>>>>>>>
\begin{align} %(30)
\label{var_fun_I}
\ket{\Psi_\mathrm{I}(c)} = \ket{\psi_\mathrm{I}(1;c)} 
\otimes \ket{\psi_\mathrm{I}(2;c)} \cdots 
\otimes \ket{\psi_\mathrm{I}(N;c)} 
\end{align}
%<<<<<<<<<<<<<<<<<<<<<<<<<<<<<<<<<<<<<<<<<<<
as depicted schematically in Fig.~\ref{scs_I}(a). 
Here $\ket{\psi_\mathrm{I}(p;c)}$ is a singlet cluster 
of four $\frac{1}{2}$-spins in the $p$th unit cell; 
this is denoted by a loop filled in grey. 
Each singlet cluster is a linear combination of two valence bond 
states written as
%>>>>>>>>>>>>>>>>>>>>>>>>>>>>>>>>>>>>>>>>>>>
\begin{align} %(31)
\label{cluster_I}
\ket{\psi_\mathrm{I}(c)} = \ket{\alpha_\mathrm{I}} 
+c\ket{\beta_\mathrm{I}} , 
\end{align}
%<<<<<<<<<<<<<<<<<<<<<<<<<<<<<<<<<<<<<<<<<<<
where 
$\ket{\alpha_\mathrm{I}}$ and $\ket{\beta_\mathrm{I}}$ 
are the valence bond states defined 
in the right hand side of Fig.~\ref{scs_I}(b). 
We have abbreviated index $p$ of the unit cell for simplicity. 
The valence bond state $\ket{\alpha_\mathrm{I}}$ is the same 
as that shown within a loop in VBS~I (Fig.~\ref{vbs}(a)). 
The valence bond state $\ket{\beta_\mathrm{I}}$ contains 
a $J_2$-valence-bond. 
The coefficient $c$ should be 0 in the limit of $j\rightarrow 0$, 
and may be small for region I-A ($J_1 > J_2$). 
But it should have a substantial amplitude 
in region I-B ($J_1 < J_2$). 
For finite $c$, the two valence bond states locally resonate 
in a singlet cluster to contribute to energy gain; 
this effect is examined in subsection \ref{subsec:boundary}, 
where  the phase boundary  between Gap~I and Gap~II phases 
is discussed energetically. 

%-->-->-->-->-->-->-->-->-->-->-->-->-->-->-->-->
%Fig. 9
\begin{figure}[t]
\begin{center}\leavevmode
\includegraphics[width=0.85\linewidth]{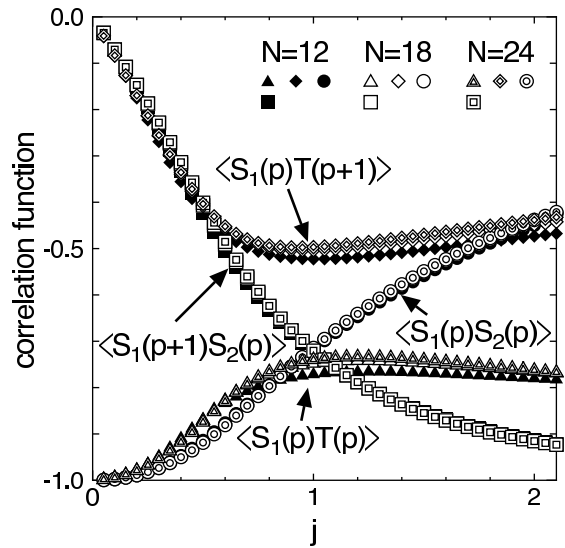}
\end{center}
\caption{Short range correlation functions 
for $r$ = 0.3 and $k$ = 0.2 calculated numerically 
for $N$ =12, 18, and 24 ($N$: the number of spins).
} 
\label{cor}
\end{figure}
%--<--<--<--<--<--<--<--<--<--<--<--<--<--<--<--<

We examined which valence bonds really contribute to 
the ground state by numerically calculating the short range 
correlation functions. 
For a typical case ($r$ = 0.3 and $k$ = 0.2), 
results are shown in Fig.~\ref{cor}. 
As $j$ increases, 
$| \aver{\v{S}_1(p) \cdot \v{S}_2(p)} |$ decreases and 
$| \aver{\v{S}_1(p+1) \cdot \v{S}_2(p)} |$ increases. 
This means that the contribution from the $J_2$-valence-bonds 
becomes large in comparison with that from the 
$J_1$-valence-bonds. 
It is also known that $| \aver{\v{S}_1(p) \cdot \v{T}(p+1)} |$ 
takes the maximum and $| \aver{\v{S}_1(p) \cdot \v{T}(p)} |$ 
takes the minimum around region I-B. 
Hence a $K_1$-valence-bond is reduced in region I-B. 
Instead, a valence bond between $\v{T}(p+1)$ and 
$\v{S}^{(1)}_1(p)$ (or $\v{S}^{(2)}_1(p)$) develops, 
although there is no exchange interaction 
between $\v{T}(p+1)$ and $\v{S}_1(p)$. 
All these results are consistent with the SCS~I picture. 

%-->-->-->-->-->-->-->-->-->-->-->-->-->-->-->-->
%Fig. 10
\begin{figure}[t]
\begin{center}\leavevmode
\includegraphics[width=0.9\linewidth]{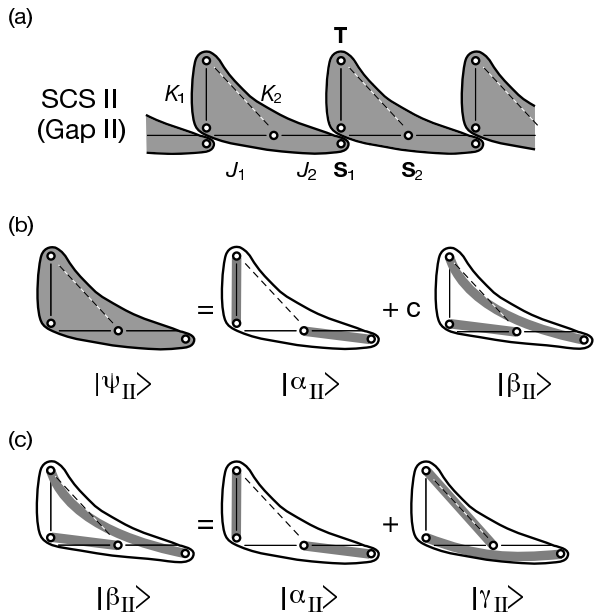}
\end{center}
\caption{(a) SCS~II, the SCS picture for Gap~II phase 
($S_1=1$, $S_2=\frac{1}{2}$, and $T=\frac{1}{2}$). 
SCS~II is a tensor product form of singlet clusters. 
(b) A singlet cluster $\ket{\psi_\mathrm{II}(c)}$ 
(the left hand side) in SCS~II. 
It is represented as a linear combination of 
two valence bond states, $\ket{\alpha_\mathrm{II}}$ and 
$\ket{\beta_\mathrm{II}}$ (the right hand side). 
$c$ is the coefficient of the linear combination. 
(c) Identity among local valence bond states. 
$\ket{\beta_\mathrm{II}}$ (the left hand side) is 
exactly the sum of 
$\ket{\alpha_\mathrm{II}}$ and $\ket{\gamma_\mathrm{II}}$ 
(the right hand side). 
}
\label{scs_II}
\end{figure}
%--<--<--<--<--<--<--<--<--<--<--<--<--<--<--<--<

%=== subsection D
\subsection{SCS picture for Gap~II phase}
\label{subsec:gap_II}

We call the SCS picture for the Gap~II phase SCS~II. 
Similarly to the case of SCS~I, 
the SCS~II state is constructed by 
assuming  the following requirements: 

(i)  The ground state is continuously modified to 
the VBS~II state without global rearrangement of 
the valence bond configuration. 

(ii)  The ground state is invariant under the translation 
by a single unit cell. 

(iii) The ground state contains a substantial amount of 
component with $J_1$-valence-bonds in region II-B. 

Under the above requirements, we take SCS~II wave function 
in the tensor product form: 
%>>>>>>>>>>>>>>>>>>>>>>>>>>>>>>>>>>>>>>>>>>>
\begin{align} %(32)
\label{var_fun_II}
\ket{\Psi_\mathrm{II}(c)} = \ket{\psi_\mathrm{II}(1;c)} 
\otimes \ket{\psi_\mathrm{II}(2;c)} \cdots 
\otimes \ket{\psi_\mathrm{II}(N;c)} 
\end{align}
%<<<<<<<<<<<<<<<<<<<<<<<<<<<<<<<<<<<<<<<<<<<
as depicted schematically in Fig.~\ref{scs_II}(a). 
Here $\ket{\psi_\mathrm{II}(p)}$ is a singlet cluster 
of four $\frac{1}{2}$-spins in the $p$th unit cell; 
this is denoted by a loop filled in grey. 
Each singlet cluster is a linear combination of two valence bond 
states written as
%>>>>>>>>>>>>>>>>>>>>>>>>>>>>>>>>>>>>>>>>>>>
\begin{align} %(33)
\label{cluster_II}
\ket{\psi_\mathrm{II}(c)} = \ket{\alpha_\mathrm{II}} 
+c\ket{\beta_\mathrm{II}} , 
\end{align}
%<<<<<<<<<<<<<<<<<<<<<<<<<<<<<<<<<<<<<<<<<<<
where 
$\ket{\alpha_\mathrm{II}}$ and $\ket{\beta_\mathrm{II}}$ 
are the two valence bond states defined 
in the right hand side of Fig.~\ref{scs_II}(b). 
We have abbreviated index $p$ of the unit cell for simplicity. 
The valence bond state $\ket{\alpha_\mathrm{II}}$ is 
the same as that shown within a loop in 
VBS~II (Fig.~\ref{vbs}(b)). 
The valence bond state $\ket{\beta_\mathrm{II}}$ contains 
a $J_1$-valence-bond. 
The coefficient $c$ should be 0 in the limit of 
$j\rightarrow \infty$, 
and may be small for region II-A ($J_1 < J_2$). 
But it should have substantial amplitude 
in region II-B ($J_1 > J_2$). 
For finite $c$, the two valence bond states locally resonate 
in a singlet cluster to contribute to energy gain; 
this effect is examined 
in subsection \ref{subsec:boundary}. 

The wave function (\ref{cluster_II}) of the singlet cluster is 
also represented as 
%>>>>>>>>>>>>>>>>>>>>>>>>>>>>>>>>>>>>>>>>>>>
\begin{align} %(34)
\label{cluster_IId}
\ket{\psi_\mathrm{II}(c)} = (1+c) \ket{\alpha_\mathrm{II}} 
+ c \ket{\gamma_\mathrm{II}} 
\end{align}
%<<<<<<<<<<<<<<<<<<<<<<<<<<<<<<<<<<<<<<<<<<<
using identity $\ket{\beta_\mathrm{II}}$ = 
$\ket{\alpha_\mathrm{II}}$ + $\ket{\gamma_\mathrm{II}}$, 
which is shown in Fig.~\ref{scs_II}(c). 
$\ket{\gamma_\mathrm{II}}$ includes the valence bond 
on the $K_2$-interaction and contributes to energy gain 
in region II-B ($K_2 \gg K_1$). 

%=== subsection E
\subsection{Boundary between Gap~I and Gap~II phases}
\label{subsec:boundary}

First, we consider the $j>1$ part of the phase boundary; 
it separates regions I-B and II-A as seen in Fig.~\ref{regions}. 
The SCS~II wave function in region II-A is close to 
VBS~II with  $J_2$-valence-bonds, 
while the SCS~I wave function in region I-B contains 
$J_2$-valence-bonds only in part 
in the linear combination (Eq.~(\ref{cluster_I})). 
Then, since the energy gain owing to $J_2$-valence-bonds 
in region II-A is always larger than that in region I-B, 
one might expect that region II-A 
would extend to the whole area of $j>1$. 
The reason why region I-B actually exists is attributed to 
the energy gain by local resonation between 
the two valence bond states within each singlet cluster 
(Eq.~(\ref{cluster_I})). 
When $k$ or $j$ increases, the SCS~I becomes less favorable 
and VBS~II becomes of advantage because of 
strong $K_1$- or $J_2$-valence-bonds. 

Second, we consider the $j<1$ part of the phase boundary; 
it separates regions II-B and I-A as seen in Fig.~\ref{regions}. 
The SCS~I wave function in region I-A is close to 
VBS~I with $J_1$-valence-bonds, 
while the SCS~II wave function in region II-B contains 
$J_1$-valence-bonds only in part 
in the linear combination (Eq.~(\ref{cluster_II})). 
Then, since the energy gain owing to $J_1$-valence-bonds 
in region I-A is always larger than that in region II-B, 
one might expect that region I-A 
would extend to the whole area of $j<1$. 
The reason why region II-B actually exists is attributed to 
the energy gain by local resonation between 
the two valence bond states within each singlet cluster 
(Eq.~(\ref{cluster_II})). 
Since the resonation is enhanced by the frustration due to 
$K_2$-interactions, the area of region I-B increases 
with increasing $r$ ($= K_2 /2K_1$) as really 
seen in Fig.~\ref{phase_nlsm} and Fig.~\ref{phase_num}. 
If $r$ is fixed, the increase of $k$ ($= K_1 /2J_1$) 
diminishes the effect of $J_1$-interactions, and 
makes the effect of resonation in SCS~II advantageous. 
This is the reason why SCS~II appears for relatively large $k$ 
in the presence of frustration $r$. 

% 8 --- variational calculation --------------------------
\section{Variational Calculation}
\label{sec:variational}

In the preceding section, we explained that the SCS picture 
represents the essence of each ground-state phase. 
However it does not guarantee 
that each SCS wave function is quantitatively satisfactory. 
In this section, we perform variational calculation using 
the SCS~I and the SCS~II wave functions 
to examine the quantitative correctness of the wave functions. 

%=== subsection A
\subsection{Variational calculation for SCS~I}
\label{subsec:var_I}

The variational wave function for SCS~I is Eq.~(\ref{var_fun_I}) 
with variational parameter $c$, which is the coefficient of 
the linear combination. 
Then the energy per unit cell in energy unit $J_1$ is written as 
%>>>>>>>>>>>>>>>>>>>>>>>>>>>>>>>>>>>>>>>>>>>
\begin{align} %(35)
\epsilon_\mathrm{I}(c) 
&= \frac{1}{NJ_1} 
\frac{\bra{\Psi_\mathrm{I}(c)} H \ket{\Psi_\mathrm{I}(c)}}
{\braket{\Psi_\mathrm{I}(c)}{\Psi_\mathrm{I}(c)}} 
\nonumber \\
&= \frac{\bra{\psi_\mathrm{I}(c)} h_\mathrm{I} \ket{\psi_\mathrm{I}(c)}}
{\braket{\psi_\mathrm{I}(c)}{\psi_\mathrm{I}(c)}} ,  
\end{align}
%<<<<<<<<<<<<<<<<<<<<<<<<<<<<<<<<<<<<<<<<<<<<
where $H$ is the total Hamiltonian (\ref{Hamiltonian}), and 
$h_\mathrm{I}$ is the reduced Hamiltonian 
for a singlet cluster in SCS~I given as 
%>>>>>>>>>>>>>>>>>>>>>>>>>>>>>>>>>>>>>>>>>>>
\begin{align} %(36)
h_\mathrm{I} 
= \v{S}_1^{(1)} \cdot \v{S}_2
+ j \v{S}_2 \cdot \v{S}_1^{(2)}
+ k \v{S}_1^{(2)} \cdot \v{T} . 
\end{align}
%<<<<<<<<<<<<<<<<<<<<<<<<<<<<<<<<<<<<<<<<<<<<

\begin{figure}[t]
\begin{center}\leavevmode
\includegraphics[width=0.88\linewidth]{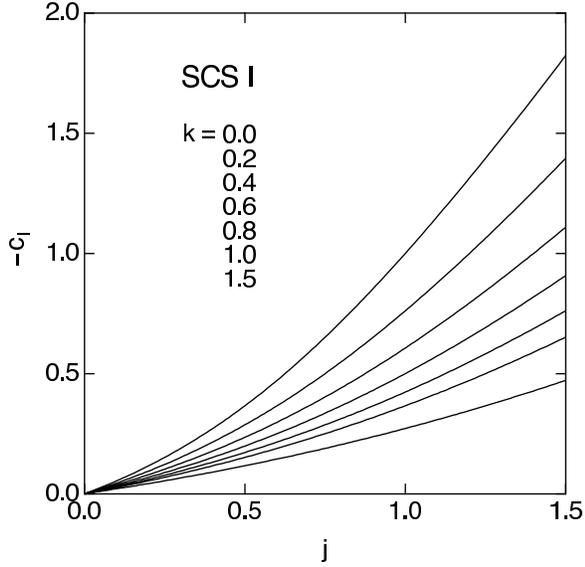}
\end{center}
\caption{ 
Coefficient $c_\mathrm{I}$ of the linear combination in the variational 
wave function for SCS~I 
($S_1=1$, $S_2=\frac{1}{2}$, and $T=\frac{1}{2}$). 
The optimal coefficient $c$ does not depend on $r$ in the 
wave function. 
The lines are in the order of ascending $k$ 
from the top to the bottom. 
} 
\label{coeff_I}
\end{figure}

After straightforward calculation, we have the following 
formula: 
%>>>>>>>>>>>>>>>>>>>>>>>>>>>>>>>>>>>>>>>>>>>
\begin{align} %(37)
\epsilon_\mathrm{I}(c) = - \frac{3}{4} 
\frac{1+k-(1+k+j)c+jc^2}{1-c+c^2} . 
\end{align}
%<<<<<<<<<<<<<<<<<<<<<<<<<<<<<<<<<<<<<<<<<<<<
Minimizing $\epsilon_\mathrm{I}(c)$ with respect to $c$, 
we have the optimal value $c_\mathrm{II}$ 
of the coefficient of the linear combination: 
%>>>>>>>>>>>>>>>>>>>>>>>>>>>>>>>>>>>>>>>>>>>
\begin{align} %(38)
\label{c_I}
c_\mathrm{I} = 
\frac{1 + k - j - \sqrt{(1+k)^2 - (1+k)j + j^2}}{1+k} . 
\end{align}
%<<<<<<<<<<<<<<<<<<<<<<<<<<<<<<<<<<<<<<<<<<<<
The result is shown in Fig.~\ref{coeff_I}, where  
the left axis represents $-c_\mathrm{I}$. 
It is independent of $r$, since a $K_2$-interaction is 
between different singlet clusters. 
In the limit of $j \rightarrow 0$, we have $c_\mathrm{I} = 0$ 
and the wave function reduces to VBS~I. 
The value $|c_\mathrm{I}|$ increases with increasing $j$, 
and for $j \gtrsim 1$ 
the term of $\ket{\beta_\mathrm{I}}$ including $J_2$-valence-bonds 
becomes dominant in $\ket{\psi_\mathrm{I}(c)}$. 
This behavior is consistent with the argument 
about SCS~I in the preceding section. 

%-->-->-->-->-->-->-->-->-->-->-->-->-->-->-->-->
%Fig. 11
%--<--<--<--<--<--<--<--<--<--<--<--<--<--<--<--<

%=== subsection B
\subsection{Variational calculation for SCS~II}
\label{subsec:var_II}

The variational wave function for SCS~II is Eq.~(\ref{var_fun_II}) 
with variational parameter $c$, which is the coefficient of 
the linear combination. 
Then the energy per unit cell in energy unit $J_1$ is written as 
%>>>>>>>>>>>>>>>>>>>>>>>>>>>>>>>>>>>>>>>>>>>
\begin{align} %(39)
\epsilon_\mathrm{II}(c) 
&= \frac{1}{NJ_1} 
\frac{\bra{\Psi_\mathrm{II}(c)} H \ket{\Psi_\mathrm{II}(c)}}
{\braket{\Psi_\mathrm{II}(c)}{\Psi_\mathrm{II}(c)}} 
\nonumber \\
&= \frac{\bra{\psi_\mathrm{II}(c)} h_\mathrm{II} \ket{\psi_\mathrm{II}(c)}}
{\braket{\psi_\mathrm{II}(c)}{\psi_\mathrm{II}(c)}} , 
\end{align}
%<<<<<<<<<<<<<<<<<<<<<<<<<<<<<<<<<<<<<<<<<<<<
where $H$ is the total Hamiltonian (\ref{Hamiltonian}), and 
$h_\mathrm{II}$ is the reduced Hamiltonian 
for a singlet cluster in SCS~II given as 
%>>>>>>>>>>>>>>>>>>>>>>>>>>>>>>>>>>>>>>>>>>>
\begin{align} %(40)
h_\mathrm{II} 
= k \v{T} \cdot \v{S}_1^{(2)}
+ \v{S}_1^{(2)} \cdot \v{S}_2
+ j \v{S}_2 \cdot \v{S}_1^{(1)}
+ 2kr \v{T} \cdot \v{S}_2 . 
\end{align}
%<<<<<<<<<<<<<<<<<<<<<<<<<<<<<<<<<<<<<<<<<<<<

%-->-->-->-->-->-->-->-->-->-->-->-->-->-->-->-->
%Fig. 12
\begin{figure}[t]
\begin{center}\leavevmode
\includegraphics[width=0.88\linewidth]{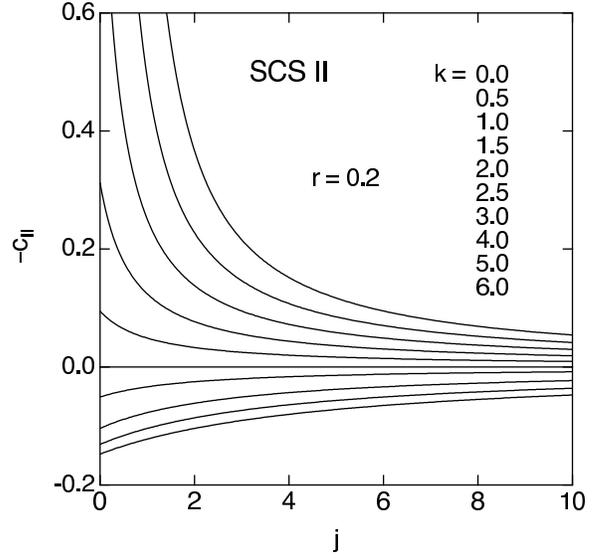}
\end{center}
\caption{ 
Coefficient $c_\mathrm{II}$ of the linear combination in the variational 
wave function for SCS~I in the case of $r=0.2$ 
($S_1=1$, $S_2=\frac{1}{2}$, and $T=\frac{1}{2}$). 
The lines are in the order of ascending $k$ 
from the top to the bottom. 
} 
\label{coeff_II}
\end{figure}
%--<--<--<--<--<--<--<--<--<--<--<--<--<--<--<--<

After straightforward calculation, we have the following 
formula: 
%>>>>>>>>>>>>>>>>>>>>>>>>>>>>>>>>>>>>>>>>>>>
\begin{align} %(41)
\epsilon_\mathrm{II}(c) = - \frac{3}{4} 
\frac{j+k-(1+k+j-2kr)c+c^2}{1-c+c^2}
\end{align}
%<<<<<<<<<<<<<<<<<<<<<<<<<<<<<<<<<<<<<<<<<<<<
Minimizing $\epsilon_\mathrm{II}(c)$ with respect to $c$, 
we have the optimal value $c_\mathrm{II}$ 
of the coefficient of the linear combination: 
%>>>>>>>>>>>>>>>>>>>>>>>>>>>>>>>>>>>>>>>>>>>
\begin{align} %(42)
\label{c_II}
c_\mathrm{II} &= \frac{a - \sqrt{a^2 +ab + b^2}}{a+b} 
\end{align}
%<<<<<<<<<<<<<<<<<<<<<<<<<<<<<<<<<<<<<<<<<<<<
with $a = j+k-1$ and $b =1-2kr$. 
The result for $r=0.2$ is shown in Fig.~\ref{coeff_II}, where  
the left axis represents $-c_\mathrm{II}$. 
In the limit of $j \rightarrow \infty$, we have $c_\mathrm{II} = 0$ 
and the wave function reduces to VBS~II. 

The result depends on $r$, since a $K_2$-interaction 
is involved in each singlet cluster. 
Except for $r$ = $1/2k$, 
$|c_\mathrm{II}|$ increases with decreasing $j$, meaning that  
$\ket{\beta_\mathrm{II}}$ including 
$J_1$-valence-bonds becomes dominant in Eq.~(\ref{cluster_II}). 
This feature is consistent with 
the SCS~II picture in the preceding section. 
In the special case of $2kr$ (= $K_2/J_1$) = 1,  
we have $c_\mathrm{II}=0$ for any values of $j$, since 
$b$ = 0 in Eq.~(\ref{c_II}). 
It is explained in the following expression for 
$\ket{\psi_\mathrm{II}(c)}$: 
%>>>>>>>>>>>>>>>>>>>>>>>>>>>>>>>>>>>>>>>>>>>
\begin{align} %(43)
\label{cluster_IIdd}
\ket{\psi_\mathrm{II}(c)} = (1+c) \ket{\beta_\mathrm{II}} 
- \ket{\gamma_\mathrm{II}} , 
\end{align}
%<<<<<<<<<<<<<<<<<<<<<<<<<<<<<<<<<<<<<<<<<<<
which is derived  from Eq.~(\ref{cluster_II}) by identity 
$\ket{\beta_\mathrm{II}}$ = 
$\ket{\alpha_\mathrm{II}}$ + $\ket{\gamma_\mathrm{II}}$ 
in Fig.~\ref{scs_II}(c). 
For $J_1$ = $K_2$, 
$\ket{\beta_\mathrm{II}}$ including a $J_1$-valence-bond 
and 
$\ket{\gamma_\mathrm{II}}$ including a $K_2$-valence-bond 
are equally weighted in Eq.~(\ref{cluster_IIdd}), 
as are expected. 

%-->-->-->-->-->-->-->-->-->-->-->-->-->-->-->-->
%Fig. 13
\begin{figure}[t]
\begin{center}\leavevmode
\includegraphics[width=0.84\linewidth]{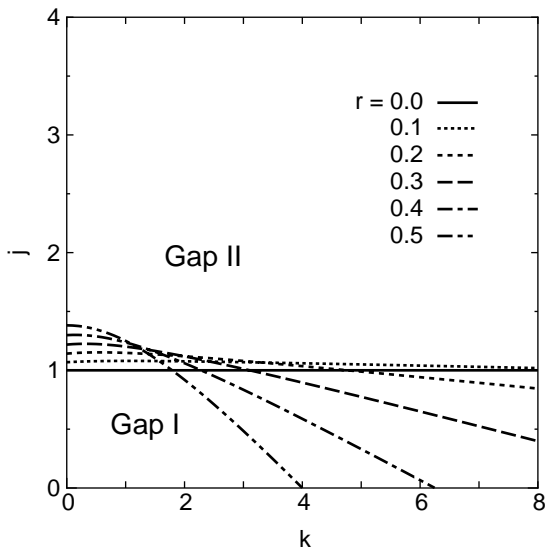}
\end{center}
\caption{ 
Phase boundaries by the variational calculation 
for the SCS wave functions
($S_1=1$, $S_2=\frac{1}{2}$, and $T=\frac{1}{2}$). 
} 
\label{phase_var}
\end{figure}
%--<--<--<--<--<--<--<--<--<--<--<--<--<--<--<--<

%=== subsection C
\subsection{Phase diagram by variational calculation}
\label{sbsec:phase_var}

In the above subsections, we have obtained 
variational energy $\epsilon_\mathrm{I}(c_\mathrm{I})$ 
of SCS~I for Gap~I phase and 
variational energy $\epsilon_\mathrm{II}(c_\mathrm{II})$ 
of SCS~II for Gap~II phase. 
The phase boundary is determined by the equation 
%>>>>>>>>>>>>>>>>>>>>>>>>>>>>>>>>>>>>>>>>>>>
\begin{align} %(44)
\epsilon_\mathrm{I}(c_\mathrm{I}) = 
\epsilon_\mathrm{II}(c_\mathrm{II}) . 
\end{align}
%<<<<<<<<<<<<<<<<<<<<<<<<<<<<<<<<<<<<<<<<<<<<
The results for various $r$ are shown in Fig.~\ref{phase_var}. 
These phase boundaries qualitatively agree with the numerical 
ones in Fig.~\ref{phase_num}, and support the correctness of 
the SCS pictures. 

Region I-B in the variational phase diagram is wider 
in $k$-direction and narrower in $j$-direction 
than the accurate one (Fig.~\ref{phase_num}). 
Further, region II-B is narrower than the accurate one. 
Remembering that regions I-B and II-B exist 
because of local resonation of valence bond states, 
inclusion of longer-range valence bonds resonating with each other 
is possibly effective to improve the variational wave functions 
in regions I-B and II-B. 
The ground states in regions I-A and II-A are close to 
VBS I and VBS II, respectively, and the effect of resonation 
is relatively small. 
Hence, the inclusion of longer longer-range valence bonds 
does not seem to be very effective to improve 
the variational wave functions. 

% 9 --- summary --------------------------
\section{Summary and Discussion}
\label{sec:summary}

In this paper, we investigated the nonmagnetic ground states of a mixed spin chain with side chains with weak frustration. So far, these kind of models have not been investigated in depth 
despite their possible rich physics. 
We have chosen the present model (\ref{Hamiltonian}) 
(Fig.~\ref{lattice}) 
as a simple and nontrivial one, which will be a good starting point. 
We examined the system in various approaches: 
a NLSM method, a numerical diagonalization method, 
an inspection of limiting cases, 
an physical interpretation based on SCS pictures, and 
a variational calculation for SCS wave functions. 

NLSM methods have been developed for simple 
spin chains without side chain. 
In the present work, we formulated a NLSM method 
for the typical spin chain with side chains. 
The NLSM method analytically provides a ground-state 
phase diagram in the $k$-$j$ parameter space 
for various values of $S_1$, $S_2$, and $T$. 
In the special case of 
$S_1$ =1, $S_2$ = $\frac{1}{2}$, and $T$ = $\frac{1}{2}$, 
the phase diagram contains two quantum disordered 
phases, Gap~I and Gap~II, in each of which 
the system has a spin-gap. 

We also examined the case of $S_1$ =1, $S_2$ = $\frac{1}{2}$ 
and $T$ = $\frac{1}{2}$ by the numerical diagonalization 
for finite chains. 
Using the method of twisted boundary condition, 
we have determined phase boundaries. 
When frustration is not strong ($r \lesssim 1$), 
there are two spin-gap phases in the $k$-$j$ parameter space. 
The numerical results confirms the qualitative correctness of 
the present NLSM method. 

The limiting cases of $j \rightarrow 0$ and 
$j \rightarrow \infty$ are precisely and analytically treated. 
For $ j \ll 1$, the Hamiltonian (\ref{Hamiltonian}) describes 
an array of weakly coupled 3-spin units. 
As $k$ increases from 0, the ground state of 
each 3-spin unit changes from singlet to triplet. 
Accordingly the whole spin chain undergoes a phase transition 
from the VBS~I state to the Haldane state. 
For $ j \gg 1$, the system is described by an effective 
Hamiltonian where no phase transition occurs with 
changing $k$ as long as $r$ is small. 
Considering the continuity to the large $k$ limit, 
the ground state is described as a state similar to 
the VBS~II state. 

There are regimes where no VBS picture explains 
the ground state for $S_1$ =1, $S_2$ = $\frac{1}{2}$ 
and $T$ = $\frac{1}{2}$. 
To explain the whole phase diagram, 
we proposed two SCS pictures; 
SCS~I and SCS~II for Gap~I and Gap~II phases, respectively. 
Each SCS is a wave function of a tensor product form 
of singlet clusters. 
A singlet cluster in both the SCS's is a local linear combination 
of two valence-bond states of two different patterns. 
The resonation contributes to the energy gain of the system, 
and the whole phases are consistently explained. 

To quantify the SCS pictures, we performed variational 
calculations with the wave functions representing 
SCS~I and SCS~II. 
The phase boundary between Gap~I and Gap~II phases are 
determined by equating the energy of 
the minimized wave function for SCS~I to that for SCS~II. 
The resultant phase diagram approximately reproduces 
the phase diagram by the numerical diagonalization. 

Thus we have obtained three phase diagrams: 
Fig.~\ref{phase_nlsm} by the NLSM method, 
Fig.~\ref{phase_num} by the numerical diagonalization, and 
Fig.~\ref{phase_var} by the variational calculation. 
The phase diagram of Fig.~\ref{phase_num} is accurate, 
since the extrapolation by finite size systems is reliable. 
The other phase diagrams qualitatively agree with 
the accurate one,  and both the methods are shown to 
be useful. 

So far, we have examined the spin chain with side chains 
when the parameter $r$ measuring frustration is not large 
($r \lesssim 1$). 
For larger $r$, wider variety of phases are expected. 
For example, 
we can extend the analysis in the limit of $j \rightarrow 0$ 
to the region of $r > 1$, i.~e. $k_c < 1$. 
For $k_c < k < 1$, effective spins of 3-spin units with 
spin magnitude 1 ferromagnetically interacts with the nearest 
neighbors and form a ferromagnetic ground state. 
In terms of the original spins, the ground state is ferrimagnetic. 
Further, by the numerical diagonalization, 
we have found various ferrimagnetic phases with different 
magnetization in the strongly frustrated regime. 
The study of these ferrimagnetic phases is in progress 
and will be reported in a separate paper. 

% ----------------------------------------
\section*{ACKNOWLEDGMENTS}

The numerical diagonalization program is based on the package 
TITPACK ver.~2 coded by H. Nishimori. 
The numerical computation in this work has been carried out 
using the facilities of the Supercomputer Center, 
Institute for Solid State Physics, University of Tokyo 
and Information Technology Center,  University of Tokyo.  
This work is partly supported by 
Fund for Project Research in Toyota Technological Institute, and 
by Innovative Research Organization, Saitama university. 

\bigskip

% ----------------------------------------


\begin{thebibliography}{99}
\bibitem{simple_chain} %[9]
The model is also considered as a spin chain 
with first neighbor couplings $K_1$, $K_2$, and $J_2$, 
and with second neighbor coupling $J_1$.
\bibitem{degree_f} %[12] 
One of $\v{L}_1$, $\v{L}_2$, and 
$\v{L}_{\bot}$ is a dummy variable. 
It turns to be absorbed by the others, and only two variables 
remain as seen in Eq.~(\ref{fluctuation_variable}). 
Hence the number of the original degrees of freedom has conserved.
\bibitem{symmetrize} %[16] 
For the equivalence, $\v{S}_1^{(1)}$ 
and $\v{S}_1^{(2)}$ are symmetrized in a wave function.
\bibitem{equiv_VBS_I} %[17] 
The wave function (\ref{sing_wf}) for a 3-spin unit 
is expressed as
$\kket{0, 0}$ = 
$( | \v{S}_1^{(1)}\v{T} \rangle | \v{S}_1^{(2)}\v{S}_2 \rangle 
+ | \v{S}_1^{(2)}\v{T} \rangle | \v{S}_1^{(1)}\v{S}_2 \rangle )
/\sqrt{2}$, 
where $|\v{S}\v{S}' \rangle$ means 
the valence bond state  between 
$\v{S}$ (=  $\v{S}_1^{(1)}$ or $\v{S}_1^{(2)}$) 
and $\v{S}'$ (= $\v{S}_2$ or $\v{T}$). 
This is a part of VBS~I in Fig. \ref{vbs}(a).
\end{thebibliography}
\end{document}